\documentclass[11pt,twoside]{article}

\usepackage{asp2006}
\usepackage{epsf}
\usepackage{psfig}
\usepackage{lscape}

\begin{document}
\title{The PennState/Toru\'n Center for Astronomy  Search for Planets around Evolved Stars.}   

\begin{quote}
Andrzej{} Niedzielski$^{1,2}$ and Aleksander Wolszczan$^{2,1}$

{\itshape $^1$Toru\'n Centre for Astronomy, Nicolaus Copernicus University,\\
ul. Gagarina 11, 87-100 Toru\'n, Poland}

{\itshape $^2$Department of Astronomy and Astrophysics, Pennsylvania State
University, 525 Davey Laboratory, University Park, PA 16802}
\end{quote}

\begin{abstract} 
We present the motivation for and the first results from a large radial velocity search for planets around red giants with the 9.2-m Hobby-Eberly Telescope.
\end{abstract}

\section*{Introduction}

Since the discovery of the first extrasolar planetary system \citep{1992Natur.355..145W} and the first planet orbiting a Sun-like star \citep{1995Natur.378..355M} more than 200 planets have been detected beyond the Solar System. 
Aside from the most successful surveys of solar type stars, a spectacular progress has been  
made in searches for planets around low-mass dwarfs, which, among other goals, have been driven
by the anticipation that Earth-mass planets can be found in their habitable zones
\citep{2005-PThPS.158...24M, 2005_PThPS.158...43M}.
However, over time, it has become
apparent that to achieve a truly satisfying level of understanding of planet formation and
evolution, the surveys have to be extended to other types of stars. 
For example,
surveys of white dwarfs, which probe ancient planetary systems -
survivors of the evolution of their parent stars - exemplify an extension of
planet searches to the endpoint of stellar evolution \citep[e.g.][]{Kepler05, 2007Natur.449..189S}.
In addition, searches for neutron star planets can
provide information on planets around massive stars \citep{1993ApJ...419L..65T}
and on planet formation in extreme, post-supernova environments \citep{2003ApJ...591L.147K, 2006Natur.440..772W}.

More than a decade ago, precision radial velocity (RV) studies have established that
GK-giant stars exhibit RV variations ranging from days to many hundreds of days
\citep[e.g.][]{walker89, HC93, 1994ApJ...422..366H}. Enough observational
evidence has been accumulated to identify three distinct sources of this variability,
namely stellar pulsations, surface activity, and a presence of substellar companions.
As Doppler searches for planets around main sequence (MS)
stars become inefficient for spectral types earlier than F6-F8, because of paucity
of spectral features and their rotational broadening, extending studies of planetary
system formation and evolution to stellar masses substantially larger than 1 M$_{\odot}$ is 
observationally difficult. A potentially very efficient, indirect way to remove this difficulty
is to conduct surveys of post-MS giants. These evolved stars have cool atmospheres and many
narrow spectral lines, which can be utilized in RV measurements to give
an adequate precision level ($<$10 m s$^{-1}$). Discoveries of planets around post-MS
giants, in numbers comparable to the current statistics of planets around MS-dwarfs
\citep[e.g.][]{2006ApJ...646..505B}, will most certainly provide the much needed 
information on planet formation around intermediate mass MS-progenitors ($\geq 1.5 M_{\odot}$)
and they will create an experimental basis with which to study 
dynamics of planetary systems orbiting evolving stars \citep[e.g.][]{1998Icar..134..303D}. 
Sufficiently large surveys of post-MS giants should furnish enough
planet detections to meaningfully address the problem of a long-term survival of
planetary systems around stars that are off the MS and
on their way to the final white dwarf stage.

\section*{Strategy and observations}

In order to address the above issues, we have joined the existing surveys 
\citep[e.g.][and references therein]{2006A&A...457..335H, sato07} 
with our own long-term project to search for planets around 
evolved stars with the 9.2-m Hobby-Eberly Telescope (HET) \citep{lwr98}
and its High Resolution Spectrograph (HRS) \citep{tull98}. The sample of stars we have been monitoring
since early 2004 is composed of two groups, approximately equal in numbers. The first one
falls in the ``clump giant'' region of the HR-diagram \citep{jim98}, which contains stars of various
masses over a range of evolutionary stages. The second group comprises stars, which have
recently left the MS and are located $\sim$1.5 mag above it. Generally, all
our targets, a total of $>$900 GK-giants brighter than $\sim$11 mag, occupy the area in
the HR-diagram, which is approximately defined by the MS, the instability strip, and the coronal
dividing line 
(a narrow strip in the HR-diagram marking the transition between stars with steady hot coronae and those with cool chromospheric winds \citep{1979ApJ...229L..27L}).
If the frequency of occurence of planets
around MS-progenitors of GK-giants is similar to that of planets around solar-type stars,
our survey should detect 50-100 planets and planetary systems, which,
together with the detections from similar projects, will provide a firm basis for studies
of planetary system formation and evolution around $>1M_{\odot}$ stars.

Observations were made    in the queue scheduled mode \citep{HetQ}. The HRS was used in the R=60,000 resolution mode with a gas cell ($I_2$) inserted into the optical path, and it
was fed with a 2 arcsec fiber.  Typically, the signal-to-noise ratio per resolution element  (at 594 nm) 
was $\sim$200 for the stellar spectra taken with the gas cell, and $\geq$250 for the templates.

The spectra consisted of 46 echelle orders recorded on the ``blue'' CCD chip (407.6-592 nm) and 24 orders on the ``red'' one (602-783.8 nm). The spectral data 
used for RV measurements were extracted from the 17 orders, which cover the 505 to 592 nm range of the $I_2$ cell spectrum.

Because the HRS spectra extend far beyond the range occupied by the $I_2$ lines, we were able to measure variations of line profiles to monitor stellar activity in the same spectra which were used for radial velocity determination. To avoid any contamination by the $I_2$ spectrum,
we have selected lines in the wavelength range redwards of 660 nm.  We have measured
two line parameters, the line bisector span and its curvature (Nowak \& Niedzielski, this volume). The "blue" high SNR template spectra, exposed without gas cell were used to obtain rotational velocities via cross-correlation with a slow rotator, (Nowak \& Niedzielski, this volume).

The observing strategy has been optimized for the HET taking into account the fact that the number of targets is large, about 2/3 of them are fainter than 8 mag, and that they are approximately randomly distributed over the sky. 
 Measurements of a particular
target star begin with 2-3 exposures, typically 3-6 months apart, to check for any RV variability
exceeding a 30-50 m s$^{-1}$ threshold. 
The interval between observations also matches the HET ability to point at a given object, usually no more than 1-2 times per night, thus favoring the expected long term RV variations.
If a significant variability is detected, the star is scheduled
for more frequent observations, and, if the RV variability is confirmed, it becomes part of
the high priority list.

The  data reduction was performed
 using  IRAF\footnote{IRAF is distributed by the National Optical Astronomy Observatories, which are operated by the Association of Universities for Research in Astronomy, Inc., under cooperative agreement with the National Science Foundation.} scripts.
Radial velocities were measured by means of the commonly used $I_2$ cell calibration technique
\citep{Butler+1996}. 
A template spectrum was constructed from a high-resolution Fourier Transform Spectrometer (FTS) $I_2$ spectrum and a high signal-to-noise
stellar spectrum measured without the $I_2$ cell. 
Doppler shifts were derived from least-square fits of template spectra to
stellar spectra with the imprinted $I_2$ absorption lines.
The resultant radial velocity measurement for each epoch was derived as a mean value of
the independent determinations from the 17 usable echelle orders. The corresponding
uncertainties of these measurements were calculated assuming that errors obeyed
the Student's t-distribution and they typically fell in a 4-5 m s$^{-1}$ range at 1$\sigma$-level.  Radial velocities were referred to the Solar System barycenter using the \citet{1980A&AS...41....1S} algorithm.

\section*{First results.}

Since the beginning of this survey, 4 years ago, over 600 stars of our sample have been observed. More than 300 of them have been observed at multiple epochs. For all stars for which enough data exist, the rms scatter in RV can be calculated. The distribution of this scatter shows a maximum between 10 and 20 m/s, which we interpret as representing the typical intrinsic RV $\it jitter$ of the red giants. Based on this result, we consider a red giant to be RV-stable, if its $\sigma$RV$\leq$40 m/s.According to this criterion, $\sim$ 55$\%$ of our sample stars can be considered as RV stable. Another $\sim$ 20 $\%$ of the program stars show RV variations with $\sigma$RV$\geq$250 m/s and we assume that they are binaries with a stellar companion.  Yet another 20 $\%$ of them show $\sigma$RV between 40 and 250 m/s and these are the stars which constitute our sample of candidates for sub-stellar companions.

We have detected a periodic RV signal in about one-half of the stars, which belong to the group of planetary companion candidates. Typically, the estimated periods are lolnger than one year and only  a very few stars show shorter periodicities. Our current approach is that only stars with the suspected sub-stellar companions, for which at least 3 consistent orbital periods have been observed, are considered for prompt publication. This is to ensure a reliable rejection of all sources of observed RV variations other than the presence of a low-mass companion. 

Our first discovery is a sub-stellar companion to the red giant HD 17092 \citep{2007arXiv0705.0935N}. 
In the absence of any correlation of the observed
360-day periodicity in radial velocities with the standard
indicators of stellar activity, the observed radial velocity variations
are most plausibly explained in terms
of a Keplerian motion of a planetary-mass body around the star. With the estimated
stellar mass of 2.3M$_\odot$, the minimum mass of the planet is 4.6M$_J$.
The planet's orbit
is characterized by a mild eccentricity of $e$=0.17 and
a semi-major axis of 1.3 AU. This is the tenth published detection of
a planetary companion around a red giant star.

\acknowledgements 
The project is 
supported in part by the Polish Ministry of Science and Higher Education grant 1P03D 007 30.
AW also acknowledges a partial support from the NASA Astrobiology Program. 
The Hobby-Eberly Telescope (HET) is a joint project of the University of Texas at Austin, the Pennsylvania State University, Stanford University, Ludwig-Maximilians-Universit\"at M\"unchen, and Georg-August-Universit\"at G\"ottingen. The HET is named in honor of its principal benefactors, William P. Hobby and Robert E. Eberly.

\end{document}